\documentstyle[aps,prl,twocolumn,epsf]{revtex}
\begin{document}

{\bf Conditions for abrupt failure in the DFBM}

We argue that the existence of abrupt failure in the democratic fiber
bundle model (DFBM) is more general than concluded by da Silveira
in his comment \cite{comment}. In this goal, we first reformulate his
equation (1)
in a much more intuitive way\,: $fx_{i+1} \equiv (F/N_0)(N_{i+1}/F) = 1 -
P(1/x_i)
\equiv 1 - P(F/N_i)$, which leads to $F/N_0 = (F/N_{i+1}) [1 - P(F/N_i)]$.
At equilibrium,
$N_{i+1}$ and $N_i$ are replaced by the number $n$ of remaining intact
fibers and
we retrieve the usual equation used in \cite{PRL1},
\begin{equation}
F/N_0 = (F/n) [1 - P(F/n)] \equiv x_n [1 - P(x_n)]~,
\label{eq1}
\end{equation}
which expresses that the total bundle will not break under a load $F$ if
there are
$n$ fibers in the bundle, each of which can withstand the stress $F/n$.
In contrast, the iterative equation (1) of
\cite{comment} is just a numerical scheme and has no physical interpretation.
It may be misinterpreted as representing
the sequence of fiber ruptures given
small increment of the applied force. 
See \cite{bursts} for proper derivations of the power law distributions
of the genuine fiber rupture burst sizes. 

The bundle breaks down when $F$ reaches the maximum of $N_0 x_n [1 - P(x_n)]$. 
In the case when $N_0 x_n [1 - P(x_n)]$ has a single maximum,
we showed \cite{PRL1} that the rate of fiber failure
diverges with a square root singularity 
on the approach towards global failure (critical behavior)
when the above function is quadratic at the maximum.
Note that the distinction between the cases (i), (ii) and (iii) in
\cite{comment} are immaterial since 
the onset of failure are qualitatively the same, all being critical.
Instead, an abrupt ``first-order'' rupture occurs when the
maximum happens to be at the minimum strength $x_{min}$. 
This condition comprises the case studied in \cite{PRL1} and
is the same as stated in
\cite{comment}, i.e., $p(x_{min})>1/x_{min}$, 
strictly equivalent
to $d\{x_n [1 - P(x_n)]\}/dx_n < 0$ at $x_n=x_{min}$, 
which follows directly from our analysis \cite{PRL1}.

We can generalize this further. Indeed, the most general condition for a brutal rupture is that
\begin{equation}
 \pm d\{x_n [1 - P(x_n)]\}/dx_n|_{x_n=x^{*\pm}} < 0~,
\end{equation} 
 i.e. that the function $x_n [1-P(x_n)]$ has
a discontinuity with a change of sign in its slope at its maximum $x^*$.
Explicitly, this gives 
\begin{equation}
p(x^{*-}) < {1 - P(x^*) \over x^*} < p(x^{*+})~,
\label{ree}
\end{equation}
i.e. the differential distribution $p(x) = {dP \over dx}$ must have a jump that is
sufficiently large at $x^*$. The previous case corresponds to the situation where the 
jump occurs at $x_{min}$ but this is a very particular case.
This condition accounts for the more subtle cases where the discontinuity
at $x_{min}$ does not tell the whole story even when $P(x)$ is monotonous.
Consider for instance a Weibull distribution  $P(x)= 1-\exp (-([x-x_{min}]/D)^m)$,  for
$x>x_{min}$, and $P(x)=0$ for $x<x_{min}$. Condition (\ref{ree}) gives 
$x^* = x_{min}$ and $m<1$.
The rupture is first-order for $m<1$, critical for $m>1$,  and for $m=1$ it is
first order for $D<x_{min}$ and critical otherwise.  
More complicated scenarios occur, because condition (\ref{ree}) only
expresses the existence of a jump, and not the fact that this is a global
jump. Another condition must ensure that $x^*$ is the global maximum
and not only a local one.
For instance, for $m=1/2$,  $x [1- P(x)]$ has 
another maximum at $x=2D(1+\sqrt{1-x_{min}/D})$ in addition to
the ``ridge'' at $x_{min}$.
This implies that the abrupt rupture at $x_{min}$ is only partial.
After that, the applied force has to increase to climb a barrier 
whose peak corresponds to a critical rupture down to $n=0$.

This class of condition (\ref{ree})
is probably relevant for real materials that do not have
a continuous distribution.
The habit to use continuous distributions such as the Weibull law
and others stems from their ability to fit rupture data of large
macroscopic systems. These fits, as
in most statistical analysis, are controlled by the regions where the data
is plentiful and not by the extreme tails. 
Without exploring the tails,
it is thus very difficult to assert
statistically whether the fiber strength distribution 
is smoothed or exhibit jumps.

As for the role of disorder,  in addition to the explicit example
discussed in \cite{PRL1},
we have also studied
the cases where $P(x)$ is a $\tanh$, a power law, and
the Weibull distribution $(m=1)$ above a minimum strength.
In these cases, the discontinuity condition at $x_{min}$ for
first-order rupture to occur
consistently translate into a requirement of small disorder which is
represented by the width of the relevant distribution.

In sum, we have refuted the claim in \cite{comment} 
that the nature of the rupture process in the DFBM depends on the
``disorder distribution only via its large $x$ behavior''.

D. Sornette$^{1,2}$ and K.-T. Leung$^3$ and J. V. Andersen$^4$, 

$^1$ LPMC, CNRS and Universit\'{e} de Nice-Sophia Antipolis, Parc Valrose,
06108 Nice, France

$^2$ ESS and IGPP, UCLA, Los Angeles, California 90095-1567

$^3$ Institute of Physics, Academia Sinica, Nankang, Taipei, Taiwan 11529,
R.O.C.

$^4$ Department of Mathematics, Imperial College, 180 Queen's Gate, London
SW7 2BZ, England


\vspace*{-0.5truecm}

\begin{references}
\vspace*{-1.6truecm}
\bibitem{comment} R. da Silveira, previous comment.

\bibitem{PRL1}  J.V. Andersen, D. Sornette and K.-T. Leung, Phys. Rev. Lett.
{\bf 78}, 2140 (1997)

\bibitem{bursts} D. Sornette, J.Phys. I France {\bf 2}, 2089 (1992);
P.C. Hemmer and A. Hansen, J. Appl. Mech., {\bf 4}, 909 (1992);
A. Hansen and P.C. Hemmer, Phys. Lett. A {\bf 184}, 394 (1994)



\end{references}
\end{document}